\documentclass[aps,prd,preprint,groupedaddress,showkeys,showpacs]{revtex4}
\usepackage{epstopdf} 
\usepackage{graphicx} \usepackage{rotating}

\def\beq{\begin{equation}}
\def\eeq{\end{equation}}
\def\bea{\begin{eqnarray}}
\def\eea{\end{eqnarray}}  

\def\eq#1{{Eq.~(\ref{#1})}}
\def\fig#1{{Fig.~\ref{#1}}}

\newcommand{\as}{\alpha_S}

\newcommand{\Lb}{\left(}
\newcommand{\Rb}{\right)}
\parskip=\itemsep

\newcommand{\A}{{\cal A}}

\def\pom{{I\!\!P}}
\begin{document}


\title{\bf \Large The Phenomenology of Pomeron Enhancement
\footnote{Based on a talk by U. Maor at LISHEP 2009, UERJ, Rio de Janeiro, 
16-23 Jan. 2009.}}
\author{E. Gotsman}
\email[]{Email: gotsman@post.tau.ac.il}
\author{E. Levin}
\email[]{Email: leving@post.tau.ac.il}
\author{U. Maor}
\email[]{Email: maor@post.tau.ac.il}
\author{J.S. Miller}
\email[]{Email: jeremymi@post.tau.ac.il}
\affiliation
{Department of Particle Physics, School of Physics and Astronomy\\
Raymond and Beverly Sackler Faculty of Exact Science\\
Tel Aviv University, Tel Aviv, 69978, Israel}
\begin{abstract} 
Multi Pomeron interactions are the main 
source of high mass diffraction. Their role in 
high energy dynamics greatly influences 
the predictions for high energy cross sections and 
survival probabilities of hard diffraction channels,
notably, diffractive Higgs production at the LHC.
Our approach, is motivated by the fact that we obtain 
a very small value  for
the fitted slope of the Pomeron trajectory, 
which justifies the use of perturbative QCD for soft scattering. 
Our suggested model differs from the proposal of 
the Durham KMR group which is based on a parton model interpretation 
of the Reggeon calculus in the complex J-plane in which multi Pomeron 
vertices are arbitrarily defined.  
The theoretical input and predictions of the two groups, 
as well as their data analysis and procedures 
are compared and evaluated.
\end{abstract}
\maketitle
\section{Introduction}
The purpose of this communication is to discuss the role of 
multi Pomeron interactions in high energy scattering. 
Our study is two fold: on the one hand,
we wish to assess the relevance of 
multi Pomeron interactions in the calculation of soft scattering cross
sections in the ISR-Tevatron range, for which data is available.
Our main goal, though, is to seek 
conclusive evidence for multi Pomeron
interactions at the LHC and Auger.
The study we  present is essential for realistic 
estimates of inelastic hard diffraction rates, in particular,  
central exclusive diffractive Higgs production at the LHC, as it is
necessary to have a reliable calculation of the     
gap survival probability for this process \cite{heralhc}.
In the following we discuss and assess the latest versions of 
two classes of eikonal models: 
\newline 
1) The Tel Aviv (GLMM) group has two publications: GLMM(08)\cite{GLMM1} 
in which our model is presented, detailing our calculation of 
the enhanced Pomeron diagrams. 
The theoretical basis of this model is further explored in 
GLMM(09)\cite{GLMM2}. 
\newline
2) The  Durham (KMR) group  has presented three recent versions 
of its model: KMR(07)\cite{RMK1}, KMR(08)\cite{RMK2} 
and LKMR(09)\cite{LKMR}. 
KMR(08) results, spread over six recent publications,  
are based on KMR(07) with a more detailed parametrization 
of the Pomeron contribution.  
LKMR(09)\cite{LKMR} is a more modest model in which high mass diffraction 
originates exclusively from the leading triple Pomeron diagram 
with secondary Regge corrections. 
Although the Tel Aviv and Durham models have a similar
philosophy, their formulations, data analysis
and predictions differ significantly.
In the following we shall discuss the consequences of
these differences. 
\section{Good-Walker eikonal models}
\par
Current eikonal models are multi channel, 
including both elastic and diffractive
re-scatterings of the initial projectiles\cite{heralhc}.
This is a consequence of the Good-Walker (GW) mechanism\cite{GW} in which
the proton (anti-proton) wave function has elastic and diffractive
components. Models based on the GW 
mechanism reproduce the total and elastic cross sections 
well, but fail to describe the diffractive cross section data 
(see Refs.\cite{GLMM1,LKMR,GLM}). 
Theoretically\cite{AK}, these deficiencies can be eliminated by the 
introduction of multi Pomeron interactions 
leading to high mass diffraction. These "Pomeron-enhanced" contributions, 
are derived from Gribov's Reggeon calculus\cite{VG}. 
The zero order, on which these calculations are based, 
is Mueller's triple Pomeron high mass SD formalism\cite{Mueller}. 
\par
Consider\cite{GLM} a vertex with an incoming hadron $|h\rangle$ 
and outgoing diffractive 
system approximated as a single state $|D\rangle$. 
The Good-Walker mechanism is based on the observation that 
these states do not diagonalize the $2{\times}2$ interaction matrix. 
We denote the interaction matrix eigenstates by $\psi_1$ and $\psi_2$. 
The wave functions of the incoming hadron and outgoing diffractive state are 
\beq \label{2CHh}
\psi_h=\alpha\,\psi_1+\beta\,\psi_2\,,
\eeq
\beq \label{2CHD}
\psi_D=-\beta\,\psi_1+\alpha \,\psi_2\,,
\eeq
where, $\alpha^2+\beta^2=1$.
For each of the four independent elastic scattering amplitudes
$A^S_{i,k}(s,b)$
we write its elastic unitarity equation 
\beq \label{UNIT}
Im\,A^S_{i,k}(s,b)=|A^S_{i,k}(s,b)|^2+G^{in}_{i,k}(s,b),
\eeq
in which
\beq \label{Aik}
A_{i,k}^S(s,b)=i\left(1-\exp(-\frac{1}{2}\Omega_{i,k}^S(s,b))\right),
\eeq
\beq \label{Gik}
G_{i,k}^{in}(s,b)=(1-\exp(-\Omega_{i,k}^S(s,b))).
\eeq
$G^{in}_{i,k}$ is the summed probability for
all non GW induced inelastic final states.
From \eq{Gik} we deduce that $P^S_{i,k}(s,b)=exp(-\Omega^S_{i,k}(s,b))$
is the probability that the GW $(i,k)$ projectiles will reach the 
final large rapidity gap (LRG) interaction in their initial state, 
regardless of their prior re-scatterings.  
\par
For $p$-$p$ and $\bar p$-$p$ scattering $A^S_{1,2}=A^S_{2,1}$,
which reduces the number of independent amplitudes to three.
The corresponding elastic, SD and DD amplitudes are 
\beq \label{EL}
a_{el}(s,b)=
i\{\alpha^4A^S_{1,1}+2\alpha^2\beta^2A^S_{1,2}+\beta^4\A^S_{2,2}\},
\eeq
\beq \label{SD}
a_{sd}(s,b)=
i\alpha\beta\{-\alpha^2A^S_{1,1}+(\alpha^2-\beta^2)A^S_{1,2}+\beta^2A^S_{2,2}\},
\eeq
\beq \label{DD}
a_{dd}=
i\alpha^2\beta^2\{A^S_{1,1}-2A^S_{1,2}+A^S_{2,2}\}. 
\eeq
For more details see Refs.\cite{heralhc,GLMM1,GLM} and references 
therein.
\par
Eikonal models based on the GW mechanism use a Regge like formalism in which 
the soft Pomeron trajectory is given by
\beq \label{Pom}
\alpha_{\pom}(t)=1+\Delta_{\pom}+\alpha_{\pom}^{\prime}t.
\eeq
The corresponding opacity is
\beq \label{Omega}
\Omega^S_{i,k}(s,b)=\nu^S_{i,k}(s)\Gamma^S_{i,k}(s,b,\alpha_{\pom}^{\prime}). 
\eeq
$\nu^S_{i,k}(s)=g_{i}g_{k}(\frac{s}{s_0})^{\Delta_{\pom}}$
and $\Gamma^S_{i,k}$ are the $b$-space profiles. 
The profiles are constructed so as to reproduce the differential 
cross sections. 
The normalization and constraints on the large $b$ behaviour of the
profiles, are determined from the data analysis. 
\par
In GLMM(08,09) an ($i,k$) $b$-profile is given as
the $b$-transform of a two pole $t$-profile ($t=-q^2$). 
Setting $\alpha_{\pom}^{\prime}$= 0, the profiles are energy independent, 
\bea \label{prof}
&&\frac{1}{(1\,\,+\,\,q^2/m^2_i)^2}\times \frac{1}{(1\,\,+\,\,q^2/m^2_k)^2}
\,\,\,\Longrightarrow\,\,\,
\Gamma \Lb b; m_i, m_k;\alpha_\pom^{\prime} = 0 \Rb.
\label{SDB}
\eea
A small energy dependence is introduced via 
\beq \label{MS}
m^2_i \,\,\,\Longrightarrow\,\,\,m^2_i(s)\,\,\equiv\,\,\,\frac{m^2_i}{1\,\,+\,\,
\alpha'_\pom\ln(s/s_0) /4m^2_i}.
\eeq
The above parametrization is compatible with the requirements of 
analyticity/crossing symmetry at large $b$, pQCD at large $q^2$ 
and Regge at small $t$. For details see Ref.\cite{GLMM2}. 
\par 
Models in which diffraction is exclusively given by the GW
mechanism, were studied by GLMM(08) and LKMR(09). 
KMR(07,08) do not discuss a GW model on its own.
This was considered in an earlier publication\cite{KMR} denoted KMR(00).
GLMM(08) fitted Pomeron parameters for its GW model are presented in
Table~\ref{t1}.
The data available in the $S{\bar{p}}pS$-Tevatron range 
is not sufficient to constrain 
the Pomeron parameters. To overcome this problem 
both GLMM(08) and LKMR(09) include, in addition to the exchanged Pomeron, 
also secondary Regge exchanges.
This enables them to extend their models to the lower ISR energies, and
compare with the abundant  data available at these energies. 
The two groups have considered rather different data bases for their fits.
We shall assess these choices in section IV. Regardless, 
both groups, utilizing their GW models, reproduce the 
elastic sectors of their data bases remarkably well, with comparable 
$\chi^2/d.o.f.$=0.87 and 0.83 respectively. 
KMR(00), who tune rather than fit their data base, obtained compatible results.   
The three GW models considered, fail to reproduce the 
diffractive cross sections.
\par
The fitted Pomeron trajectory parameters of the above  
GW models are compatible: 
\newline 
1) The $\Delta_{\pom}$ values obtained are very similar. 
GLMM(08) and LKMR(09) fit $\Delta_{\pom}=0.12$. KMR(00) tuned
$\Delta_{\pom}=0.10$. 
\newline
2) The above GW eikonal models find very small $\alpha_{\pom}^{\prime}$ values. 
GLMM(08) fitted $\alpha^{\prime}_{\pom}=0.012$, 
while LKMR(09) fit is $\alpha^{\prime}_{\pom}=0.033$. 
KMR(00) have a somewhat higher $\alpha^{\prime}_{\pom}=0.066$.  
\newline
3) A good reproduction of
$\frac{d\sigma_{el}}{dt}$, where $t \leq 0.5 GeV^2$,
has been attained by all GW models we have discussed.
As we shall see in the next section, the same good reproduction of 
$\frac{d\sigma_{el}}{dt}$ is also obtained  in GW+$\pom$-enhanced models. 
We shall discuss in what follows the consequences of this observation. 
\begin{table}
\begin{tabular}{|l|l|l|l|l|l|l|l|l|}
\hline
&$\Delta_\pom $ & $\,\,\,\beta$ &\,\,\,\,\,\,\, $\alpha^{\prime}_{\pom}$&
\,\,\,\,\,\,\, $g_1$ &\,\,\,\,\,\,\,  $g_2$ &
\,\,\,\,\,\,\,$m_1$ &\,\,\,\,\,\,\,$m_2$ & $\chi^2/d.o.f.$ \\
\hline
\,\,\,\,\,\,\,\,\,\,\,\,\,\,\,GW\,\,&0.120 & 0.46 & 0.012 $GeV^{-2}$ & 
1.27 $GeV^{-1}$ & 3.33 $GeV^{-1}$ & 0.913 $GeV$& 0.98 $GeV$ & 
\,\,\,\,\,\,0.87 \\
\hline
\,GW+$\pom$-enhanced  & 0.335 & 0.34 &  0.010 $GeV^{-2}$& 5.82 $GeV^{-1}$ &
239.6 $GeV^{-1}$ & 1.54 $GeV$& 3.06 $GeV$ & 
\,\,\,\,\,\,1.00 \\
\hline
\end{tabular}
\caption{Fitted parameters for GLMM(08) GW and GW+$\pom$-enhanced models.}
\label{t1}
\end{table}
\section{Multi Pomeron Interactions}
\par
The triple Regge diagram, leading to high mass soft diffraction, 
was introduced\cite{Mueller} almost 40 years ago.
CDF analysis\cite{CDF} suggests that $g_{3{\pom}}$, 
the triple Pomeron coupling, is reasonably large. 
Once we assume that 
$g_{3{\pom}}$ is not negligible, we also need to 
consider more complicated configurations 
of multi Pomeron interactions. This is the basis for 
the construction of the Pomeron-enhanced contributions 
which are consistent with t-channel unitarity. 
Note that these, non GW contributions, are contained in $G^{in}_{i,k}$,
rather than within the GW $A^S_{i,k}$ amplitudes. As such, these 
calculations have to take into account the relevant 
unitarity suppressions expressed 
in terms of the corresponding survival probabilities.
This study is of fundamental importance for a   
theoretical understanding of the Pomeron, hopefully leading to more 
precise predictions of the asymptotic behaviour of the scattering 
amplitudes. 
\par
KMR(07,08) approach is based on Ref.\cite{AK},  
which is derived from the Reggeon field calculus\cite{VG}  
with a strong emphasis on its parton model interpretation.
LKMR(09) is a much simpler model where Pomeron enhancement
is reduced to the zero order triple Regge approximation\cite{Mueller}
for SD. DD is not discussed in this model. 
As a consequence, the tuned values of $\Delta_{\pom}$ 
in the two models are very different, see Table~\ref{pom}. 
In a Regge approach, such as the above, the soft Pomeron is a simple J-pole
and the calculations of its interactions are non perturbative.
The hard Pomeron is a branch cut in the J-plane which is treated perturbatively.
Note that, in such a scheme, the hard 
Pomeron couplings are not necessarily factorisable.
Hard Pomeron coupling factorization is assumed, though, 
in most eikonal models dealing with hard diffraction.
\par
KMR(07,08) enhanced Pomeron formalism is based on 
two {\it ad hoc} assumptions: 
\newline
1) The point coupling of a multi Pomeron vertex,
$n{\pom}\,\rightarrow\,m{\pom}$, is 
\beq \label{nmP}
g_m^n\,=\,\frac{1}{2}\,g_N\,\,nm\,\lambda^{n+m-2}\,
=\,\frac{1}{2}\,nm\,g_{3\pom}\,\lambda^{n+m-3}.  
\eeq
In this notation $g_{3\pom}\,=\,\lambda g_N$, where  
$\lambda$ is a free parameter, $n+m\,>\,2$.
\newline
2) The triple Pomeron coupling strength, is assumed
to be independent of the identity (soft or hard)
of the 3 coupled Pomerons. 
\newline
The above two input assumptions lack any theoretical proof   
(see also Ref.\cite{GLMM2}). 
As such, their validity depends on strong support 
from the accompanying data analysis. 
Our assessment is that the support 
provided to this end by KMR data analysis 
is inadequate (see next section).
\par
The key input observation leading to the  GLMM(08) model 
is the exceedingly small fitted value of $\alpha^{\prime}_{\pom}$ 
obtained in our GW model fit and maintained  
in our advanced fit based on the GW+${\pom}$-enhanced model, 
see Table~\ref{t1}. 
The microscopic sub structure of the Pomeron 
is provided by Gribov's partonic interpretation of Regge theory\cite{VG}, 
in which the slope of the Pomeron trajectory is related to
$< p_t>$, the mean transverse momentum of the partons (gluons) 
associated with the exchanged Pomeron, 
$\alpha^{\prime}_\pom\,\propto\,1/<p_t>^2$. 
Our fitted $\alpha^{\prime}_\pom\,=0.010$ leads to 
our estimate that, the typical parton momentum is large. 
Regardless of its intuitive appeal, the parton model is not suitable to
describe gluonic interactions as it presumes a short range interaction
in rapidity space, while the exchange of gluon dipoles in QCD is long
range in rapidity. Recall that in pQCD the BFKL Pomeron
slope approaches zero at high enough energies as
$\alpha^{\prime}_\pom\,\propto\,1/Q_s$.
It follows that the running QCD coupling
$\as\,\propto\,\pi/\ln\Lb <p^2_t>/\Lambda^2_{QCD}\Rb \,\ll\,1$,
and we can consider it as
our small parameter, when applying pQCD estimates to
the $\pom-\pom$ interaction vertices. 
\par
Our pQCD motivated calculations are based on the MPSI approach\cite{MPSI}. 
Whereas KMR(07,08) assume a non zero $g_m^n$ point coupling of 
$n\pom \rightarrow m\pom$, GLMM(08) reduce this 
transition to the sum of all triple Pomeron fan diagrams 
contributing to this configuration. In this 
context, partons are colourless dipoles. 
This construction depends
on the basic one dipole splitting into two dipoles 
and the merging of two dipoles into one dipole, to which we assign the  
probabilities $P(1 \rightarrow 2)$ and $P(2 \rightarrow 1)$. 
For details see Refs.\cite{GLMM1,GLMM2}. 
\par
A comparison between the Pomeron trajectory parameters obtained in 
GLMM(08), KMR(07), KMR(08) and LKMR(09) is presented in Table~\ref{pom}. 
Our determination of ${\alpha^{\prime}_\pom}$ is compatible with 
the KMR(07) assumption that ${\alpha^{\prime}_\pom}=0$ 
and LKMR(09) and KMR(08) fits.  
The high $\Delta_{\pom}$ 
values presented in Table~\ref{pom} are considerably higher than 
conventional GW soft $\Delta_{\pom}$, which are smaller than 0.15.
These high values are 
compatible with BFKL and HERA high $Q^2$ (hard) DIS measurements.
Note that there is a coupling between the fitted smallness of
$\alpha^{\prime}_{\pom}$ and the large fitted value of $\Delta_{\pom}$.
The shrinkage of the forward elastic peak is a well established experimental
feature. In a traditional Regge models, the shrinkage is initiated by
the relatively large $\alpha^{\prime}_{\pom}\,=\,0.25 GeV^{-2}$. Since
a very small $\alpha^{\prime}_{\pom}$ is implied by all
GW+${\pom}$-enhanced eikonal models,
the shrinkage in these models is initiated by
the unitarity screening resulting from a large $\Delta_{\pom}$.
\begin{table}
\begin{tabular}{|l|l|l|l|l|}
\hline
&GLMM(08)&KMR(07)&KMR(08)&LKMR(09)\\
\hline
\,$\Delta_{\pom}$\,&\,\,\,\,\,\,\,\,0.335\,&\,\,\,\,\,\,\,\,0.55\,
&\,\,\,\,\,\,\,\,0.30\,&\,\,\,\,\,\,\,\,0.12\,\\
\,$\alpha^{\prime}_{\pom}$\,&\,\,\,\,\,\,\,\,0.010\,&
\,\,\,\,\,\,\,\,\,\,\,\,0\,&
\,\,\,\,\,\,\,\,0.05\,&\,\,\,\,\,\,\,\,0.033\,\\
\hline
\end{tabular}
\caption{Pomeron trajectory parameters in $\pom$-enhanced models.} 
\label{pom}
\end{table}
\par
Our pQCD Pomeron treatment is applicable to both soft and hard Pomerons
and our bare triple Pomeron coupling is, thus, universal. As such, 
our self consistent theoretical approach  supports 
the KMR(07) {\it ad hoc} assumption on the universality of the bare
$g_{3\pom}$. As noted, the bare triple Pomeron coupling 
becomes smaller with energy, due to the
monotonic decrease with increasing energy of the associated  
survival probability\cite{GLMG3,KMRG3}. 
\par
\begin{figure}
\includegraphics[width=165mm,height=70mm]{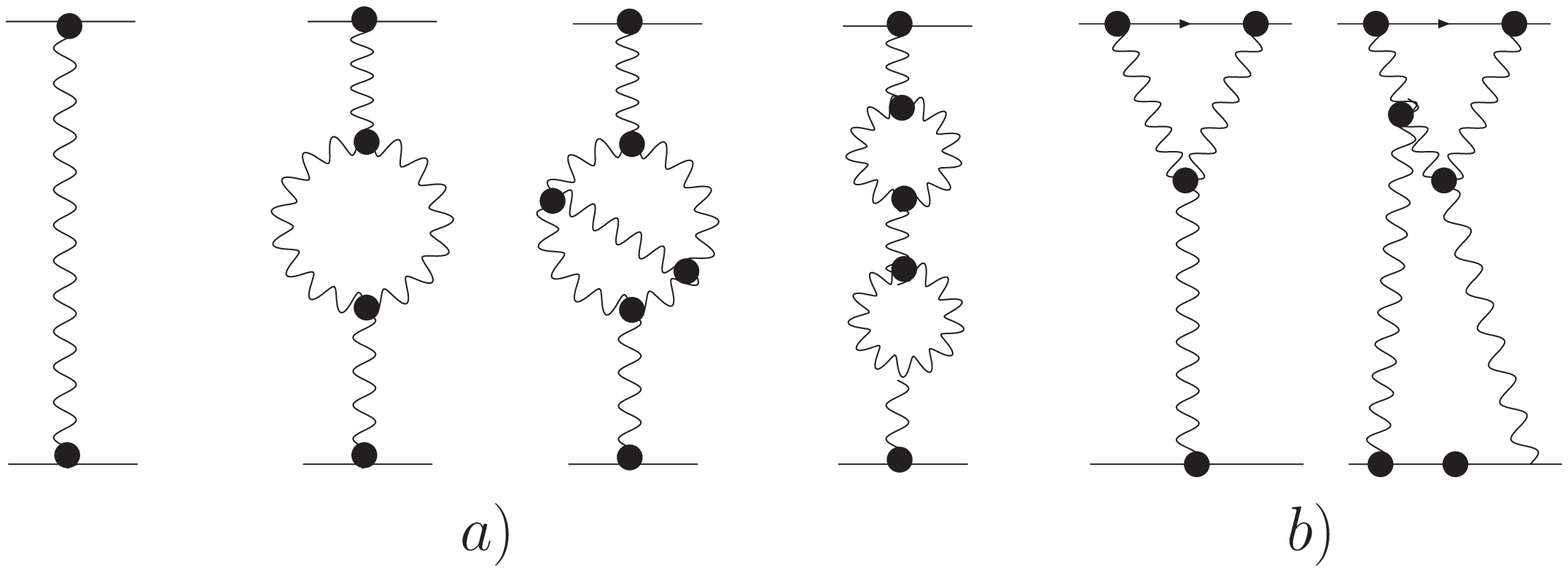}
\caption{Typical low order terms of the Pomeron Green's function. 
Enhanced Pomeron diagrams are shown in \fig{enh}a, whereas 
\fig{enh}b shows semi-enhanced diagrams which are not included 
in our calculations as yet.}
\label{enh}
\end{figure}

The high $\Delta_{\pom}$ values obtained in GLMM(08) and KMR(07,08) 
induce an energy dependent renormalization 
of $\Delta_{\pom}$, due to Pomeron loop corrections to the input Pomeron
propagator. The net result is that the effective $\Delta_{\pom}$
is reduced with increasing energy. A schematic description 
of the corresponding Pomeron Green's function  
is given in \fig{enh}. 
In GLMM(08,09) and this communication we have taken into account 
only the effect of the enhanced diagrams (\fig{enh}a), and have ignored
the semi-enhanced diagrams (\fig{enh}b). 
A more complete calculation, summing over both enhanced and semi-enhanced 
diagram is in progress. 
\par
Technically, our calculations were executed assuming that
$\alpha^{\prime}_\pom\,=0$.
This approximation implies a high energy bound of our model 
validity, at $W\,=\,10^5 GeV$, which is reached  
when $\alpha^{\prime}_\pom ln(s)$ becomes significant.
A similar bound was defined, also, in KMR(07,08).
An additional validity bound has to be introduced 
so as to control the reduction of $\Delta_{\pom}$ with energy. 
In general, this control is provided 
through higher order multi Pomeron point like vertices which constrain 
$\Delta_{\pom}$ from becoming negative. 
Since our model lacks these higher order vertices, 
its validity has to be bound. From a practical point of view, 
checking Table~\ref{DE} in the next section, we conclude that 
this validity bound is higher than the bound implied by $\alpha^{\prime}_\pom$ 
and, hence, can be neglected.  
\section{The Interplay Between Theory and Data Analysis} 
\par
There is a significant difference between the data analysis 
carried out by the Tel Aviv and Durham groups. 
This is reflected in the choice of data bases made by the 
two groups and their procedure for determining parameters, whether  
be it by fitting, tuning or assuming. 
The starting point of both investigations is the 
observation that a GW model reproduces the elastic data well, but,
its reproduction of the diffractive sector is deficient. 
Both GLMM(08), KMR(07,08) and LKMR(09) achieve an improved  
reproduction of their respective data bases, once the 
contributions of multi Pomeron diagrams are included. 
\par
\begin{figure}
\includegraphics[width=120mm,height=100mm]{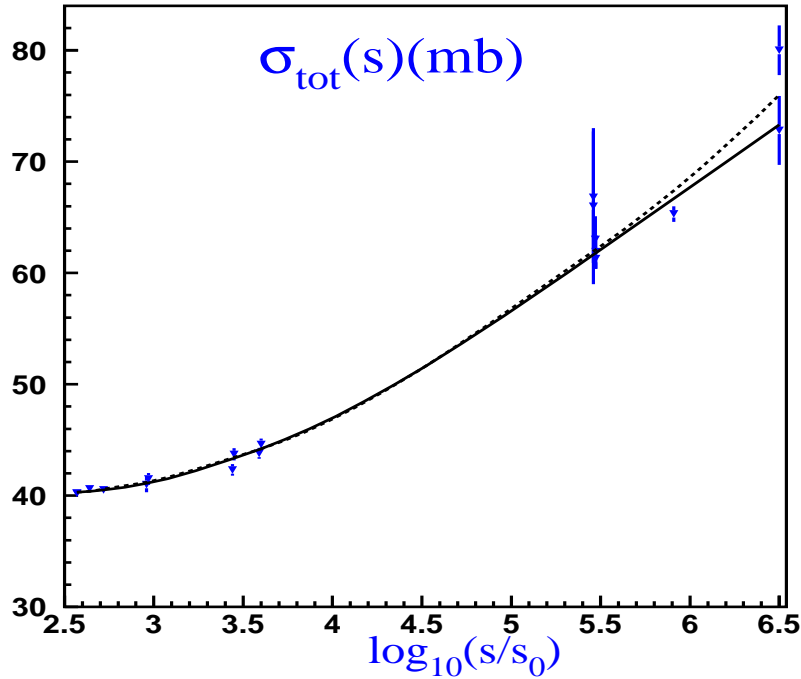}
\caption{Energy dependence of $\sigma_{tot}$ in GLMM(08) models.
Solid line corresponds to GW+${\pom}$-enhanced and dashed
line to exclusive GW.}
\label{TT}
\end{figure}

 
The data analysis of GLMM(08) is based on a diversified data base 
so as to investigate simultaneously the various theoretical 
input elements of their model.   
The fitted 55 data points include 
the total, elastic, SD and DD soft cross sections and the elastic 
forward slope in the ISR-Tevatron range, to which we have added 
a consistency check of our predicted SD forward slope, 
CDF\cite{CDF} differential elastic cross sections 
as well as its SD mass distribution at $t=0.05 GeV^2$. 
The fitting of our data base was done independently twice. 
In the first phase for a GW model and then, again, for GW+$\pom$-enhanced. 
Checking Table~\ref{t1}, we observe that, with the exception of 
$\alpha_{\pom}^{\prime}$ which is very stable, 
the fitted Pomeron parameters of the two GLMM(08) models are different. 
The dramatic changes are in $\Delta_{\pom}$ and $g_2$, which 
become much larger in the GW+$\pom$-enhanced model. 
The critical observation is that, regardless of the parameters change,
GLMM(08) second phase reproduction
of the elastic sector is as good as its first.
An example is shown in \fig{TT}.
\par
The conceptual approach of Durham group is completely different.
Their data base contains just the measured values of
$d\sigma_{el}/dt$, $\sigma_{tot}$ 
and $d\sigma_{sd}/dtd(M^2/s)$ at $t=0.05 GeV^2$. 
LKMR(09), like KMR(00), is a two phase analysis in which the 
GW variables are adjusted from the elastic data 
and frozen to be utilized without a change 
in the next phase in which the triple Regge coupling is  
determined from the SD experimental mass distributions.
KMR(07,08) do not present a GW data analysis, so we presume 
that theirs is a one phase GW+$\pom$-enhanced data analysis. 
$\alpha^{\prime}_{\pom}=0$ is assumed in KMR(07)  
and tuned in KMR(08). 
$\Delta_{\pom}$ is tuned in KMR(07,08).  
In our opinion, Durham data base is too limited to  
substantiate  their phenomenological goals.   
This may explain the shortcomings of the procedures they have 
adopted in their data analysis.   
Specifically:
\newline
1) As we saw, the features of $d\sigma_{el}/dt$
are well reproduced by all 6 models discussed in this 
paper, regardless of their presumed dynamics or specific parameters.
\newline
2) As demonstrated by GLMM(08), the refitted GW parameters 
in the GW+$\pom$-enhanced model are significantly different 
(with the exception of $\alpha^{\prime}_{\pom}$)
from those obtained in the GW fit. 
KMR models have a very simplified parametrization of the GW
proton wave components. The three components differ just in their 
effective radii which are adjusted . 
This is marginally compatible with the output
of GLMM(08) first phase fit, while its second phase fit requires an
increase of $g_2$ by a factor of 60.
\newline
3) KMR(07) reproduction of 
CDF $d\sigma_{sd}/dtd(M^2/s)$ distributions\cite{CDF} 
supposedly provides the support
for their particular introduction of multi Pomeron interactions 
and the consequent tuning of $\Delta_{\pom}$. 
We find this analysis inconclusive due to theoretical 
as well as experimental ambiguities.    
The extensive LKMR(09) analysis indicates the importance 
of triple couplings with secondary Regge contributions 
all through, including the highest, ISR-Tevatron energy range.
This implies a significant Regge background coupled to 
the triple Pomeron contribution which is the key 
element of Pomeron enhancement. 
The less extensive analysis of KMR(07) and GLMM(08)
for $d\sigma_{sd}/dtd(M^2/s)$, depends  
on the introduction of arbitrary background terms.
Moreover, LKMR(09) noted that fitting the CDF data requires a relative
rescaling of
25$\%$ between the 540 and 1800 GeV normalizations\cite{LKMR}. 
\newline
4) Ref.\cite{RMK1} claims a successful reproduction of the CDF 
high mass DD cross sections at 1800 and 630 GeV. 
We consider this claim not well established since 
the quoted data have  errors of 27$\%$ and 32$\%$ respectively.
\newline
We conclude that given its meager data base, the Durham
analysis does not have the resolution to determine
the critical Pomeron parameters.
\par
GLMM(08), KMR(07) and KMR(08) high energy Tevatron, LHC and Cosmic Rays
predicted cross sections, as well as the survival probabilities 
for exclusive central diffractive Higgs production,  
are summarized in Table~\ref{T}. We have also added to the table the 
KMR(08) predictions for the total, elastic and SD cross sections, as well 
the available partial information given on the exclusive Higgs survival
probability.
\begin{table}
\begin{tabular}{|l|l|l|l|}
\hline
& \,\,\,\,\,\,\,\,\,\,\,\,\,\,\,\,\,\,\,Tevatron
& \,\,\,\,\,\,\,\,\,\,\,\,\,\,\,\,\,\,\,\,\,\, LHC
& \,\,\,\,\,\,\,\,\,\,\,\, W=$10^5$ GeV  \\
& GLMM\,\,KMR(07)\,\,KMR(08)
& GLMM\,\,KMR(07)\,\,KMR(08)
& GLMM\,\,KMR(07)\,\,KMR(08) \\
\hline
$\sigma_{tot}$(mb)
&\,\, 73.3 \,\,\,\,\,\,\,\,\,\,74.0\,\,\,\,\,\,\,\,\,\,\,\,\,\,73.7
& \,\,\,\,92.1 
\,\,\,\,\,\,\,\,\,\,\,88.0\,\,\,\,\,\,\,\,\,\,\,\,\,\,\,91.7
&\,108.0 \,\,\,\,\,\,\,\,\,\,\,98.0\,\,\,\,\,\,\,\,\,\,\,\,\,108.0 \\
\hline
$\sigma_{el}$(mb)
& \,\, 16.3 \,\,\,\,\,\,\,\,\,\,16.3\,\,\,\,\,\,\,\,\,\,\,\,\,16.4
& \,\,\,\,20.9 
\,\,\,\,\,\,\,\,\,\,\,20.1\,\,\,\,\,\,\,\,\,\,\,\,\,\,\,21.5
& \,\,\,\,24.0 \,\,\,\,\,\,\,\,\,
22.9\,\,\,\,\,\,\,\,\,\,\,\,\,\,\,\,26.2 \\
\hline
$\sigma_{sd}$(mb)
& \,\,\,\,\, 9.8 \,\,\,\,\,\,\,\,\,\,10.9\,\,\,\,\,\,\,\,\,\,\,\,\,13.8
& \,\,\,\,11.8 
\,\,\,\,\,\,\,\,\,\,\,13.3\,\,\,\,\,\,\,\,\,\,\,\,\,\,\,19.0
& \,\,\,\,14.4  
\,\,\,\,\,\,\,\,\,\,\,15.7\,\,\,\,\,\,\,\,\,\,\,\,\,\,\,\,24.2 \\
$\sigma^{\mbox{low M}}_{sd}$
& \,\,\,\,\, 8.6 
\,\,\,\,\,\,\,\,\,\,\,\,\,4.4\,\,\,\,\,\,\,\,\,\,\,\,\,\,\,\,4.1
& \,\,\,\,10.5 \,\,\,\,\,\,\,\,\,\,\,\,\,\,5.1
\,\,\,\,\,\,\,\,\,\,\,\,\,\,\,\,4.9
& \,\,\,\,12.2\,\,\,\, 
\,\,\,\,\,\,\,\,\,\,5.7\,\,\,\,\,\,\,\,\,\,\,\,\,\,\,\,\,\,\,5.6 \\
$\sigma^{\mbox{high  M}}_{sd}$
& \,\,\,\,\, 1.2
\,\,\,\,\,\,\,\,\,\,\,\,\,6.5\,\,\,\,\,\,\,\,\,\,\,\,\,\,\,\,9.7
& \,\,\,\,\,\,\,1.3 
\,\,\,\,\,\,\,\,\,\,\,\,\,\,8.2\,\,\,\,\,\,\,\,\,\,\,\,\,\,\,14.1
& \,\,\,\,\,\,\,2.2
\,\,\,\,\,\,\,\,\,\,\,10.0\,\,\,\,\,\,\,\,\,\,\,\,\,\,\,\,18.6 \\
\hline
$\sigma_{dd}$(mb) & \,\,\,\,\, 5.4\,\,\,\,\,\,\,\,\,\,\,\,\,\,\,7.2 
& \,\,\,\,\,\,\,6.1 \,\,\,\,\,\,\,\,\,\,\,13.4
& \,\,\,\,\,\,\,6.3\,\,\,\,\,\,\,\,\,\,\,\,\,17.3 \\
\hline
$\frac{\sigma_{el}+\sigma_{diff}}{\sigma_{tot}}$
& 
\,\,\,\,\, 0.43\,\,\,\,\,\,\,\,\,\,\,0.46
& 
\,\,\,\,\,\,\,0.42\,\,\,\,\,\,\,\,\,\,\,\,0.53
& 
\,\,\,\,\,\,\,0.41\,\,\,\,\,\,\,\,\,\,\,\,\,0.57\\
\hline
$S^2_{2ch}(\%)$ &
\,\,\,\,\, 3.2  \,\,\,\,\,\,\,\,\,1.8-4.8 &
\,\,\,\,\,\,\,2.35  \,\,\,\,\,\,1.2-3.2 &
\,\,\,\,\,\,\,2.0  \,\,\,\,\,\,\,\,\,\,\,0.9-2.5 \\
\hline
$S^2_{enh}(\%)$ & \,\,\, 28.5 \,\,\,\,\,\,\,\,\,\,\,\,100 
&\,\,\,\,\,\,\,6.3  \,\,\,\,\,\,\,\,\,\,\,\,100 
\,\,\,\,\,\,\,\,\,\,\,\,\,\,\,33.3&
\,\,\,\,\,\,\,3.3 \,\,\,\,\,\,\,\, \,\,\,\,100 \\ \hline
$S^2(\%)$ &
\,\,\,\,\, 0.91  \,\,\,\,\,\,\,2.7-4.8
& \,\,\,\,\,\,\,0.15 \,\,\,\,\,\,\,1.2-3.2 \,\,\,\,\,\,\,\,\,\,\,\,1.5
&\,\,\,\,\,\,\,0.066\,\,\,\,\,\,\,0.9-2.5\\
\hline
\end{tabular}
\caption{Comparison of GLMM, KMR(07) and KMR(08) outputs.}
\label{T}
\end{table}
\par
The elastic and total cross section predictions of the quoted models
are roughly compatible and, above the Tevatron energy,
they are significantly lower than the cross sections 
obtained in models with no multi Pomeron contributions.
This is a consequence of $\Delta_{\pom}$ renormalization.
We illustrate these features in Table~\ref{DE} where we present 
the renormalized effective values of 
$\Delta_{\pom}$ in GLMM(08) and KMR(07,08) models as deduced from 
the corresponding cross section predictions for the Tevatron,
LHC and W=$10^5$ GeV.
Note that the values of $\Delta_{\pom}^{eff}$ deduced from $\sigma_{tot}$
and $\sigma_{el}$ differ, as these cross sections are screened
differently.
As is evident from Table~\ref{DE}, this behaviour persists
up to $W = 10^5 GeV$, which is the limit of validity of
both the GLMM(08,09) and the KMR(07,08) calculations. 
\par
GLMM(08) and KMR(07) $\sigma_{sd}$ predictions are reasonably compatible.
Note, though, that the two models
low and high mass diffraction contributions are inconsistent.
The KMR(07) high mass diffraction grows much faster with energy
than that predicted by GLMM(08).
On the other hand, the GLMM(08) predictions for GW diffraction
are about twice as large as those predicted by KMR(07).
GLMM(08) identifies low diffractive mass with GW diffraction with no
rapidity cut, whereas KMR(07,08) differentiate
between low and high mass diffraction by a ${\Delta }y=3$ cut.
The GLMM(08,09) and the KMR(07,08) modeling of multi Pomeron interactions
are fundamentally different. The large difference between their LHC
diffractive predictions may serve, thus, as an effective test for the
validity of the two concepts suggested in these papers. We recall that KMR(08),
which is a more advanced 3 component Pomeron model, has an even larger
SD high mass diffractive cross section.
\begin{table}
\begin{tabular}{|l|l|l|}
\hline
& \,\,Tevatron $\rightarrow$ LHC \,\,
& \,\,LHC $\rightarrow$ W=$10^5$ GeV \\
& \,\,GLMM\,\,\,\,\,\,\,KMR(07)\,\,\,\,\,\,KMR(08)
& \,\,GLMM\,\,\,\,\,\,\,KMR(07)\,\,\,\,\,\,KMR(08) \\
\hline
\,\,\,\,$\sigma_{tot}$\,\,\,\,
& \,\,\,\,\,\,0.056\,\,\,\,\,\,\,\,\,\,\,\,\,0.042
\,\,\,\,\,\,\,\,\,\,\,\,\,\,\,\,\,0.053
& 
\,\,\,\,\,0.041\,\,\,\,\,\,\,\,\,\,\,\,\,\,\,0.027
\,\,\,\,\,\,\,\,\,\,\,\,\,\,0.042 
\\
\hline
\,\,\,\,$\sigma_{el}$\,\,\,\,
& 
\,\,\,\,\,\,0.030\,\,\,\,\,\,\,\,\,\,\,\,\,0.026
\,\,\,\,\,\,\,\,\,\,\,\,\,\,\,\,\,0.033
& \,\,\,\,\,0.018\,\,\,\,\,\,\,\,\,\,\,\,\,\,\,0.017
\,\,\,\,\,\,\,\,\,\,\,\,\,\,0.025 \\
\hline
\end{tabular}
\caption{$\Delta_{\pom}^{eff}$ values obtained from $\sigma_{tot}$
and $\sigma_{el}$ predictions of GLM(08)M and KMR(07,08) models.}
\label{DE}
\end{table}
\section{Survival Probabilities}
\par
The experimental program at the LHC is focused, to a considerable extent, 
on the discovery of the Higgs boson. We shall confine our discussion to 
a Standard Model Higgs with a relatively low  
mass of 120-180 $GeV$, produced in an exclusive central diffraction,   
\beq \label{H}
p\,+p\,\rightarrow\,p\,+\,LRG\,+\,H\,+\,LRG\,+\,p.
\eeq
The advantage of this channel is that it has a distinctive signature of two 
large rapidity gaps and a favorable signal to background ratio, which is 
improved when the forward protons are tagged.   
\par
\begin{figure}
\includegraphics[width=150mm]{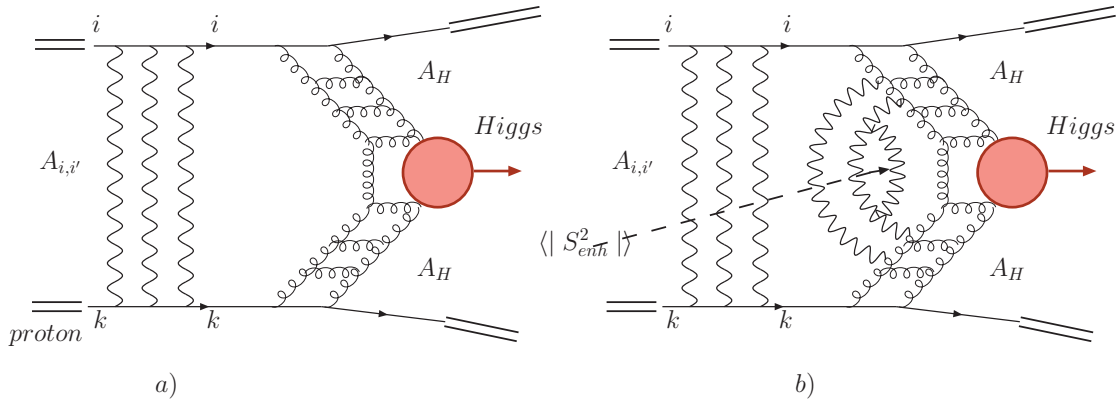}
\caption{Survival probability for exclusive central diffractive
production of the Higgs boson.
\fig{sp-dia}a shows the contribution to the
survival probability from the GW two channel component.
\fig{sp-dia}b illustrates an example of 
the additional factor $S_{enh}^2$
initiated by multi Pomeron interactions.}
\label{sp-dia}
\end{figure}

The hard pQCD calculation of this 
cross section\cite{KMRH} is reduced due to s-channel unitarity 
suppressing factors, caused by the 
re-scatterings of the initial projectiles.
We also consider the additional suppression, 
associated with t-channel unitarity, 
which is initiated by the extra 
screening induced by multi Pomeron interactions.   
The overall cross section reduction is expressed
in terms of the gap survival probability which has two components 
shown schematically in \fig{sp-dia}, 
\beq \label{S2}
S_H^2 \,=\,S_{2ch}^2\,\times \,S_{enh}^2.
\eeq 
\par
\begin{figure}
\includegraphics[width=120mm,height=100mm]{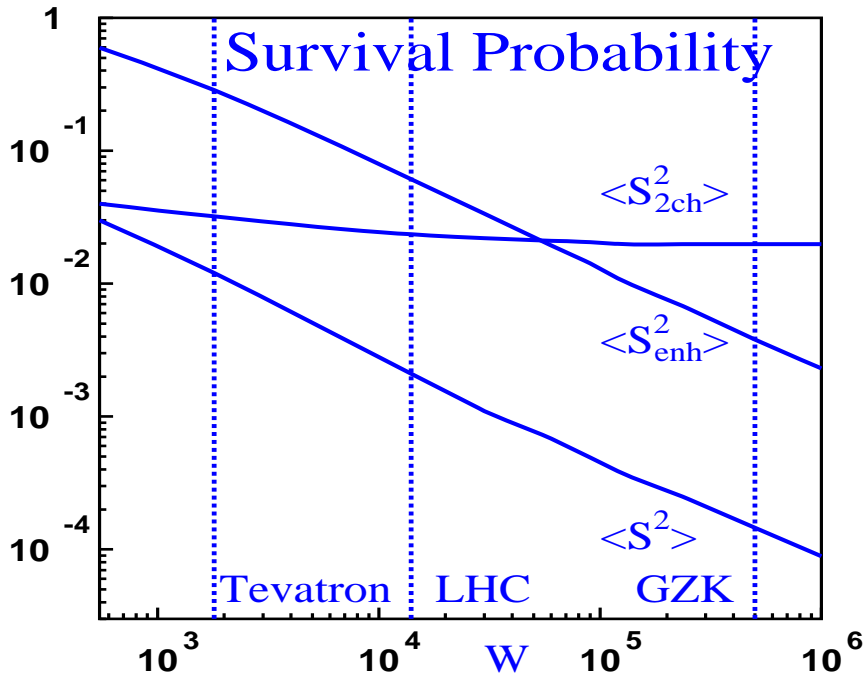}
\caption{Energy dependence of centrally produced Higgs survival probability.}
\label{sp}
\end{figure}

A detailed account of GLMM(08) $S_{H}^2$ calculations is given in 
Ref.\cite{GLMM1}. 
The Tevatron, LHC and $W = 10^5 GeV^2$ values of the above survival probabilities
are summarized in Table~\ref{T}.  
They are presented in \fig{sp} as a function of W over a wide energy range.
As can be seen, the energy dependence of $S^2_{2ch}$ is very mild. 
This is compatible with the low values
of the renormalized $\Delta_{\pom}$. 
Our $S_{2ch}^2$ calculated results in the Tevatron-LHC energy range
are compatible with KMR(07). 
Note, however, that $S_{2ch}^2$ decrease with energy in KMR(07) estimates is 
somewhat faster than in GLMM(08). Hence, the difference in the reported
results at $W = 10^5 GeV$ (see Table~\ref{T}).
There is a large discrepancy between $S_{enh}^2$ 
estimates given by GLMM(08) and KMR(07,08). 
$S^2_{enh}$ in our model 
is small and has a steep energy dependence. 
KMR(07) do not consider this suppression, i.e. they have $S_{enh}^2=1$.
In KMR(08), $S_{enh}^2=1/3$. The net result is that there is a  
large difference between GLMM(08) and
KMR(07,08) estimates of $S^2_H$. In our calculations the decrease of $S^2_H$ 
from the Tevatron to LHC is by a factor of 5-6, whereas KMR(07,08) show a 
much milder decrease of approximately a factor of 2. 
\par
In order to understand the difference 
between GLMM(08) and  
KMR(07,08) estimates of $S^2_{enh}$,  
we refer to the schematic \fig{sp-dia}.
Assume we neglect $S^2_{enh}$ (i.e., $S^2_{enh}$=1). In this case 
we have a "soft-hard" factorization. "Soft" relates to the soft 
re-scatterings of the incoming projectiles shown in \fig{sp-dia}a.
"Hard" relates to the hard diffractive process. 
$S^2_{enh}$ originates from three different, 
interfering, contributions: 
\newline
1) Multi Pomeron interactions in which Pomerons from 
the "soft" sector interact
with Pomerons from the "hard" sector. 
These interactions break the "soft-hard" factorization. 
They are included in GLMM(08,09) and KMR(08) 
calculations, but not in KMR(07) which neglected $S^2_{enh}$. 
KMR(08) claim that this suppression is relatively mild, at most a reduction  
by a factor of 3.
\newline
2) Multi Pomeron interactions confined to Pomerons in the hard 
diffractive sector (see \fig{sp-dia}b). These interactions maintain the 
"soft-hard" factorization. They are included in GLMM(08,09) but not in
either KMR(07) or KMR(08) estimates. In our assessment this suppression 
is significant. 
\newline
3) Semi-enhanced Pomeron interactions are not included in either GLMM(08) 
or KMR(08) $S^2_{enh}$ calculations.  
GLMM(09) has a calculation of $S^2_{enh}$ in which only the 
semi-enhanced diagrams are included, and find its implied suppression 
to be significant. We are planning to finish soon a comprehensive calculation 
taking into account the complete set of our model's Pomerom diagrams. 
\section{Exceedingly High Energy Behaviour}
\par
The basic GW amplitudes of the GLM models
are $A^S_{1,1}$, $A^S_{1,2}$ and $A^S_{2,2}$, 
with $b$ dependences specified in \eq{Aik} and \eq{Omega}.
These are the building blocks
with which we construct $a_{el}$, $a_{sd}$
and $a_{dd}$ (\eq{EL}-\eq{DD}).
The $A^S_{i,k}$ amplitudes are bounded by the s-channel unitarity black
disc bound of unity.
$a_{el}(s,b)$ reaches this bound at a given $(s,b)$ 
when, and only when,
$A^S_{1,1}(s,b)=A^S_{1,2}(s,b)=A^S_{2,2}(s,b)=1$, 
independent of the value of $\beta$. 
Consequently, when $a_{el}(s,b)=1$, $a_{sd}(s,b)=a_{dd}(s,b)=0$.
\par
Checking the GLMM(08) GW fitted parameters, presented in Table~\ref{t1},
we observe that $g_1$ and $g_2$, are comparable.  
Indeed, the approach of $a_{el}(s,b=0)$ to unity in GLMM(08) 
first phase GW model analysis,    
is compatible with the results obtained by KMR(07). 
This picture changes radically in GLMM(08) GW+$\pom$-enhanced model
in which the re-fitted $g_2>>g_1$ are a by-product  
of the successful reproduction of the diffractive data base.
Similar results were also obtained 
in a previous GW type GLM model\cite{GLM07}, where we were able to 
reproduce the diffractive data only after adjusting $g_2>>g_1$.
As a consequence of the above, the three basic GW amplitudes 
reach the black disc bound at different energies. 
As $g_2$ is so much larger than $g_1$,   
$A^S_{2,2}(s,b=0)$ reaches unity at a very low energy, 
$A^S_{1,2}(s,b=0)$ reaches unity at approximately W=100 GeV and  
$A^S_{1,1}(s,b=0)$ reaches unity at exceedingly high energies.  
The net result is a very slow approach of $a_{el}(s,b=0)$ toward the 
black disc bound in the GLMM(08) model, reaching the bound well above 
the LHC energy. 
Recall, that the adjusted values of $A^S_{i,k}$ are determined by a fit to 
a GW+$\pom$-enhanced model. Namely, a model based on 
both s and t-channel unitarity considerations.
\par 
The behaviour of the ratio
$R_{D} = \frac{(\sigma_{el} + \sigma_{sd} + \sigma_{dd})}{\sigma_{tot}}$
conveys information regarding the onset of
$s$-unitarity constraints at very high energies. 
Assuming that diffraction originates exclusively from the GW mechanism. 
We obtain, then, that the  
Pumplin bound\cite{Pumplin} $R_{D} \leq 0.5$. 
The non GW multi Pomeron induced diffractive contributions 
are not included in this bound 
since they originate from $G^{in}_{i,k}$.
Hence, their non screened amplitudes are suppressed by 
the survival probabilities which decrease with energy.  
The delicate balance between the increase with energy of the non screened
diffractive amplitudes, and the monotonic decrease with energy of the 
survival probabilities, is model dependent. 
Indeed, the balance between these two contributions in GLMM(08) 
and KMR(07,08) is different. 
In GLMM(08), $R_{D} < 0.5$, decreasing very slowly with energy. 
In KMR(07), $R_{D}>0.5$, increasing slowly with
energy up to W=$10^5$ GeV, which is the high energy
limit of validity for both KMR(07,08) and GLMM(08,09) models. 
The origin of the KMR(07) prediction originates from the relatively high,
and fast growing, diffractive high mass cross sections coupled to 
a minimal decrease with energy of their $S^2_{enh}$, 
which also includes their Pomeron semi-enhanced contributions.   
Judging from its fast SD increase with energy, one would expect 
the growth of $R_{D}$ in KMR(08) to be even faster. We cannot
check this expectation since KMR(08) provide only the low DD mass
prediction of their model.
Measurement of SD and DD cross sections at the LHC will provide crucial 
information on this issue. 
\section{CONCLUSIONS}
\par
We conclude with our main observations: 
\newline
1) Introducing multi Pomeron interactions, 
in addition to the conventional GW mechanism, 
in up-to-date eikonal models enables a reproduction of the 
elastic and the diffractive soft sectors.
\newline
2) We have emphasised the importance of constructing a suitable data base
to test the theoretical models and determine their free parameters. In our
opinion the Durham group data base is too small to reliably determine  
the Pomeron parameters.
\newline
3) The recent GLMM(08,09) and KMR(07,08) are contrasting models with
different diagram summations.
The novelty of the GLMM approach is that it correlates the smallness of 
the fitted $\alpha_{\pom}^{\prime}$ with the hardness of 
the presumed "soft Pomeron".
This allows one to treat the "soft Pomeron" perturbatively.
This is very different from the KMR(07,08) approach which is based on
the Reggeon calculus to which KMR have added 
two unsubstantiated assumptions relating to the multi Pomeron couplings.
\newline
4) In GLMM(08) and KMR(07,08)  the reproductions of the
elastic high energy sectors are similar. 
Pomeron enhancement, regardless of its formulation, implies a
renormalization of $\Delta_{\pom}$. We expect, thus, the total and elastic 
cross sections at the LHC and Auger energies to be smaller than the
non unitarised predictions.
\newline
5) There are severe differences between GLMM(08) and KMR(07)  
diffractive high mass predictions. 
The disagreement between GLMM(08) and KMR(08) is even larger. 
This inconsistency will probably be settled 
once these cross sections will be measured at the LHC. 
\newline
6) The GLMM(08) and KMR(08) estimates of $S^2_{enh}$ are very different.  
Both sets of calculations are not sufficiently comprehensive  
and need to be improved. 
\newline
7) In our opinion it is misleading to assume that a model which  
reproduces some data or an important variable, such as 
a gap survival probability, at the Tevatron necessarily provides LHC 
predictions which are reliable. 
The market is full of considerably different models which 
claim to reproduce the Tevatron data and its deduced variables which result, 
never the less, with a very wide and, sometimes, 
contradicting spectra of LHC predictions. Rephrasing it simply, a satisfactory  
comprehensive reproduction of the Tevatron data 
is necessary, but not sufficient to support a given model!
\newline
8) As we noted, $\Delta_{\pom}$ is renormalized by Pomeron-enhanced dynamics.
Consequently, the exceedingly high energy behaviour of $a_{el}$, $a_{sd}$
and $a_{dd}$ is determined jointly by $s$ and $t$ unitarity considerations.
The precise asymptotic behaviour of the GW amplitudes depends not only on 
$g_{3\pom}$ but also on higher multi Pomeron point couplings which are not 
included in the GLMM(08,09) approximations. 
These only become important at energies
higher than $10^5$ Gev (the validity bound of the model).
\newline
\\{\bf Acknowledgements:} 
This research was supported in part by the Israel Science Foundation, 
founded by the Israeli Academy of Science and Humanities, 
by BSF grant $\#$ 20004019 and by a grant from Israel Ministry of Science, 
Culture and Sport and the Foundation for Basic Research of the Russian Federation.

\end{document}